\documentclass[aps,12pt,tightenlines]{revtex4-1}
\usepackage{hyperref,epsfig,amssymb,amsmath,graphicx,color}

\begin{document}

\title{
Surplus Angle and Sign-flipped Coulomb Force
in Projectable Ho\v{r}ava-Lifshitz Gravity}
\author{Hang Bae Kim}\email{hbkim@hanyang.ac.kr}
\address{Department of Physics and Research Institute for Natural Sciences,
Hanyang University, Seoul 133-791, Korea}
\address{Korea Institute for Advanced Study (KIAS), Seoul 130-722, Korea}
\author{Yoonbai Kim}\email{yoonbai@skku.edu}
\address{Department of Physics, BK21 Physics Research Division,
and Institute of Basic Science,\\ Sungkyunkwan University, Suwon 440-746, Korea}

\begin{abstract}
We obtain the static spherically symmetric vacuum solutions of
Ho\v{r}ava-Lifshitz gravity theory,
imposing the detailed balance condition only in the UV limit.
We find the solutions in two different coordinate systems,
the Painlev\'e-Gullstrand coordinates and the Poincar\'{e} coordinates,
to examine the consequences of imposing the projectability condition.
The solutions in two coordinate systems are distinct
due to the non-relativistic nature of the HL gravity.
In the Painlev\'e-Gullstrand coordinates
compiling with the projectability condition,
the solution
involves an additional integration constant
which yields surplus angle and
implies attractive Coulomb force between same charges.
\end{abstract}
\pacs{98.80.Cq}
\keywords{Horava-Lifshitz gravity, projectability, surplus angle}
\maketitle

\newpage

\section{Introduction}

The UV complete theory of gravity has been a long-felt want
since general relativity (GR) and quantum field theory were established.
For past two decades, string theory has been considered to be a strongest
and possibly unique candidate.
Recently Ho\v{r}ava proposed a more practical/economical way to achieve this
goal based on quantum field theory by giving up the general covariance of
GR \cite{Horava:2009uw,Horava:2010zj}.
In this theory, the time and the space show different scaling behaviors
at the UV regime.  It modifies the high momentum behavior of the graviton
propagator and renders the theory power-counting renormalizable.

This novel feature of Ho\v{r}ava-Lifshitz (HL) gravity attracted much
interest and various aspects of the theory have been investigated,
including the properties and the consistency of the theory
\cite{Visser:2009fg,Charmousis:2009tc,Li:2009bg,Kim:2009zn},
the properties of classical solutions including black holes and their
thermodynamic properties
\cite{Lu:2009em,Nastase:2009nk},
the cosmological aspects \cite{Calcagni:2009ar,Kiritsis:2009sh},
the perturbation spectrum and the gravitational wave production
\cite{Takahashi:2009wc},
and the phenomenological sides
\cite{Sotiriou:2009gy}.
One notable feature of the theory is that the general covariance (full
4D diffeomorphism) is reduced to the foliation-preserving diffeomorphism.
Thus, we have additional degree of freedom, a dynamical scalar
\cite{Li:2009bg,Kim:2009zn},
but with a first-order (in time) equation of motion.
It gives rise to subtleties in canonical quantization, instability problem,
and strong coupling problem \cite{Charmousis:2009tc}.
On the other hand, the theory must be reduced to GR
in the IR limit.  It means that the symmetry must be enlarged at IR,
the so-called emergent symmetry.
Since HL gravity is supposed to be a quantum theory,
it depends on the running of coupling constants,
of which we do not have full understanding yet.
This may lead to phenomenological difficulties, for example,
the energy dependence in the limiting speed $c^2(E)$ and $\delta c^2(E)$.
To get the better control over these problems,
we may impose the projectability condition on the metric and
the detailed balance conditions on the coupling constants of the theory.
So far, it seems that imposing the detailed balance condition up to the IR
regime is problematic
and imposing the projectability is favorable.

Though the explicit quantization and the proof of the renormalizability are
yet to be explored, it is interesting to see the consequences of such
an approach at the IR regime, that is, in the classical solutions
and to find the observable signatures through them.
In this paper, we obtain the static spherically symmetric vacuum solutions of
HL gravity imposing the detailed balance conditions only in the UV limit.
Thus specific relations between parameters in the four or less spatial
derivative terms are not imposed.
For HL gravity, the choice of coordinates is important because the theory
lacks the general covariance and is intrinsically non-relativistic.
To define a theory, we must specify a frame where the action is defined.
Without knowledge about the preferred frame a priori,
we choose two coordinate systems, the Painlev\'{e}-Gullstrand (PG) coordinates
and the Poincar\'{e} coordinates for comparison.
The former coordinate system is better motivated for HL gravity
since it satisfies the projectability condition \cite{Horava:2009uw}.
The comparison of the solutions in two coordinate system will contrast
the differences between projectable and non-projectable HL gravity
and shed some light on the consequences of imposing the projectability condition.
Specifically, for the obtained solutions without matter distribution
in this paper, the former shows a long range effect like a surplus or deficit
solid angle irrespective of the quartic derivative terms but the latter
only has a short distance correction like the change of event horizons.
In the presence of quartic derivative terms, the effect of those in the former
is comparable to the electrostatic field of a point charge but, in some
parameter range,
the square of it has a negative value.

This paper is organized as follows.
In Section \ref{section-HL-review}
we review HL gravity, setting up our notations.
The static spherically symmetric vacuum solutions are obtained
in the PG coordinates in Section \ref{section-solution-PG}, and
in the Poincar\'{e} coordinates in Section \ref{section-solution-S}.
We conclude in Section \ref{section-conclusion}.

\section{Ho\v{r}ava-Lifshitz Gravity}
\label{section-HL-review}

The HL gravity has the invariance under the foliation-preserving diffeomorphism
$t\rightarrow\tilde{t}(t)$, $x^i \rightarrow \tilde{x}^i(t,{\bf x})$,
and a scaling behavior in the UV limit
$t\rightarrow \ell^3\,t$, $x^i\rightarrow\ell\,x^i$.
The action for the HL gravity is best described using the following
dynamical variables: the lapse function $N$, the shift functions $N_i$,
and the three-dimensional spatial metric $g_{ij}$,
with which the metric takes the ADM form
\begin{equation}
\label{ADM}
ds^{2} = -N^{2}dt^{2} + g_{ij}\left(dx^{i} + N^{i}dt\right)
     \left(dx^{j} + N^{j}dt\right),
\end{equation}
where $N^{i}\equiv g^{ij}N_{j}$.

In the UV regime there are 5 independent sixth order spatial derivative terms
saturating the $z=3$ anisotropic scaling~\cite{Sotiriou:2009bx},
\begin{align}
R^3,\quad R\square R,\quad R_{ij}R^{ij}R,\quad R_{ij}\square R^{ij},\quad
R_{ij}{R^{i}}_kR^{jk},
\end{align}
which make the theory less predictable.
Here $R_{ij}$ is the Ricci tensor of $g_{ij}$, $R=g^{ij}R_{ij}$,
and square($\square$) denotes the Laplacian with respect to the spatial
metric $g_{ij}$.
Since the detailed balance condition spoils the IR dynamics of HL gravity
and the physical motivation for its introduction is not manifest yet,
we impose it only in the vicinity of UV fixed point. Then the UV dynamics
in the limit of $z=3$ scaling is
assumed to be governed by the quadratic time derivative terms and
square of the Cotton tensor~\cite{Horava:2009uw},
\begin{equation}
C^{ij} \equiv \frac{\epsilon^{ikl}g^{jm}}{\sqrt{g}}
\left(R_{lm}-\frac{1}{4}g_{lm}R\right)_{;k},
\end{equation}
where $\epsilon^{ijk}$ is the antisymmetric tensor density with $\epsilon^{123} = 1$
and semicolon(;) denotes the spatial covariant derivative.
Then the HL gravity action under consideration is given by
\begin{eqnarray}
S_{{\rm HL}} &=& \int dt d^3x N \sqrt{g} \Bigg[
\alpha\left(K_{ij}K^{ij} - \lambda K^{2}\right) + \xi R + \sigma
\nonumber\\&&\hspace{28mm}{}
+ \beta C_{ij}C^{ij} + \gamma\frac{\epsilon^{ijk}g^{lm}}{\sqrt{g}}R_{il}R_{km;j}
+ \zeta R_{ij}R^{ij} + \eta R^{2}
\Bigg],
\label{eq-HL-action}
\end{eqnarray}
where $K_{ij}$ is the extrinsic curvature
\begin{equation}
K_{ij} \equiv \frac{1}{2N}\left(\frac{\partial g_{ij}}{\partial t}
    -N_{i;j}-N_{j;i}\right).
\end{equation}
The action \eqref{eq-HL-action} possesses 8 parameters,
$\alpha, \; \lambda,\;  \xi, \; \sigma ,\; \beta,\; \gamma,\; \zeta,\; \eta$,
and, in the IR limit, $\alpha$, $\xi$, and $\sigma$ terms dominate over higher
derivative terms. To recover GR, the renormalization group (RG) running
toward the IR limit must lead to
\begin{equation}
\lambda=1, \quad
\alpha=\frac{1}{16\pi Gc}, \quad
\xi=\frac{c}{16\pi G}, \quad
\sigma=-\frac{c\Lambda}{8\pi G},
\end{equation}
where $c$ is the speed of light, $G$ is the Newton's constant,
and $\Lambda$ is the cosmological constant.

\section{Static spherically symmetric solution in the Painlev\'e-Gullstrand coordinates}
\label{section-solution-PG}

The projectability condition states that the lapse function is a function
of the time only, that is, $N=N(t)$ in the metric (\ref{ADM}).
Then the time reparametrization, a symmetry transformation in HL gravity,
always allows the fixation of $N=1$.
The difficulties without imposing the projectability condition have
already been discussed for quantization of HL gravity~\cite{Li:2009bg},
and then we study the so-called projectable HL gravity in this section. This
projectable version is a different theory from non-projectable HL gravity
since the foliation-preserving diffeomorphism cannot turn one into the other.

To get the static spherically symmetric vacuum solution
under the projetability condition,
let us consider the static spherically symmetric metric in the PG coordinates
\begin{equation}
\label{metric-sss-pg}
ds^2 = -dt^2+\frac{1}{f(r)}[dr+n(r)dt]^2+r^2(d\theta^2+\sin^2\theta d\phi^2),
\end{equation}
which satisfies the projectability condition.
The lapse function and the shift function for the metric \eqref{ADM} are
unity and $n/f$, respectively.

The action \eqref{eq-HL-action} in terms of $f$ and $n$ is given by
\begin{align}
S_{{\rm HL}} =&\ 4\pi\int dt\,dr\, \frac{r^2}{\sqrt{f}} \left\{\;
-\alpha n^2\left[
    \frac{(\lambda-1)}{4}\left(\frac{2n'}{n}-\frac{f'}{f}\right)^2
  + \frac{2\lambda}{r}\left(\frac{2n'}{n}-\frac{f'}{f}\right)
  + \frac{2(2\lambda-1)}{r^2}
\right]
\right.\nonumber\\ &\left.{} \hspace{-9mm}
+\frac{(3\zeta+8\eta)}{2r^2}{f'}^2
+\frac{2(\zeta+4\eta)}{r^3}f'(f-1)
+\frac{2(\zeta+2\eta)}{r^4}(f-1)^2
-\frac{2\xi}{r^2}(rf'+f-1)
+\sigma
\vphantom{\left\{\left(\frac18\right)^2\right\}}
\;\right\},
\end{align}
and the action for the matter-field is
\begin{equation}
S_{{\rm M}}=4\pi\int dt\,dr\, \frac{r^2}{\sqrt{f}}\,
{\cal L}_{{\rm M}}(n,f,\Phi ),
\end{equation}
where $\Phi$ stands for all the matter fields of our consideration.
Since all the components of Cotton tensor vanish, $C^{ij}=0$ under this PG
metric \eqref{metric-sss-pg}, the contribution from the sixth and fifth order spatial derivative
terms disappears in the action and so will do in the equations of motion.
It means that the spherically symmetric solutions of our interest are
appropriate in describing up to the intermediate energy scale involving
quartic spatial derivatives
and consistent with our concern on the IR limit and the leading corrections.

The equations of motion are
\begin{eqnarray}
\label{eq-sss-pg-N}
&&
\alpha n^2\left[ (\lambda-1)r^2\left(
\frac{n''}{n}-\frac{f''}{2f}+\frac{n'^2}{2n^2}-\frac{n'f'}{nf}
+\frac{5f'^2}{8f^2}-\frac{f'}{2rf}
\right)
+2(2\lambda-1)r\frac{n'}{n}+1
\right]
\nonumber\\
&&
+(3\zeta+8\eta)\left(f''f-\frac{1}{4}f'^2\right)
-\frac{1}{r^2}\left[ (5\zeta+14\eta)(f-1)^2+2(3\zeta+8\eta)(f-1) \right]
-\xi(f-1)
\nonumber\\&&\hspace{81mm}{}
+\frac{\sigma}{2}r^2 =
r^2\left(\frac{1}{f}\frac{\partial{\cal L}_{\rm M}}{\partial f}
-\frac{1}{2}{\cal L}_{\rm M}\right),
\end{eqnarray}
and
\begin{equation}
\label{eq-sss-pg-Nr}
\alpha(\lambda-1)\left[ r^2\left(
  \frac{2n''}{n}-\frac{f''}{f}-\frac{n'}{n}\frac{f'}{f}+\frac{f'^2}{f^2}
\right) +4r\frac{n'}{n}-4 \right] + 2\alpha r\frac{f'}{f} =
\frac{1}{n}\frac{\partial{\cal L}_{\rm M}}{\partial n}.
\end{equation}
Here the equations of motion have been obtained by directly inserting
the metric ansatz \eqref{metric-sss-pg} into the action and then by
variation of the action.
Since the metric has good symmetries, \eqref{eq-sss-pg-N}--\eqref{eq-sss-pg-Nr}
coincide with the Euler-Lagrange equations for static spherically symmetric
objects, derived from the action \eqref{eq-HL-action} after assigning the
projectability condition $N=1$.

Now we consider the vacuum solution for which ${\cal L}_{\rm M}=0$.
Concerning the character of the solution, we consider the $\lambda=1$ case
for tractability, which is also consistent with the recovery of GR
in the IR limit. A newly proposed version of HL gravity, which is free from
the unwanted scalar graviton, possesses an
extra local U(1) symmetry to the foliation-preserving diffeomorphism
naturally fixes $\lambda$ to be unity~\cite{Horava:2010zj}.
With $\lambda=1$ the equation~(\ref{eq-sss-pg-Nr}) is reduced to $f'=0$
so that we write a constant solution
\begin{align}
f(r)=1+f_{0}>0.
\label{f0}
\end{align}
Inserting \eqref{f0}
into the equation (\ref{eq-sss-pg-N}), we obtain
\begin{align}
(rn^2)'= -\frac{\sigma}{2\alpha}r^{2}+\frac{\xi}{\alpha}f_{0}
+\frac{(5\zeta+14\eta)f_0^2+2(3\zeta+8\eta)f_0}{\alpha r^2}.
\label{neq}
\end{align}
The solution to this equation is
\begin{equation}
n(r)=\pm\sqrt{-\frac{\sigma}{6\alpha}r^2 + \frac{\xi}{\alpha}f_0
+\frac{r_{{\rm s}}}{r}
-\frac{(5\zeta+14\eta)f_0^2+2(3\zeta+8\eta)f_0}{\alpha r^2} }\, ,
\label{nsol}
\end{equation}
where $r_{{\rm s}}$ is the integration constant which can be identified as
$r_{{\rm s}}=2GM$ and $M$ is the mass of black hole as in the Schwarzschild
solution.
Two comments are noted before analyzing the solution.
First, if $f_0=0$, the solution reduces to that of GR
even for $\xi/\alpha\ne1$ and in the presence of the higher spatial
derivative terms.
Second,
the full set of field equations of the non-projectable version of HL gravity
up to quartic spatial derivative terms were derived in
\cite{Kiritsis:2009sh,Lu:2009em}.
The projectability condition $N=N(t)$ is not compatible with those equations
in general. Since the non-projectable version is a different theory
from the projectable version of our interest,
our analysis will be focused on the obtained solution \eqref{nsol}.

In the GR limit where $\xi/\alpha=1$ and $\sigma=\zeta=\eta=0$,
$f_0$ can be rescaled to be zero by an appropriate $t$-$r$ mixing coordinate transformation.
On the other hand, in HL gravity, such $t$-$r$ mixing coordinate transformation
is not allowed as a symmetry transformation.
Thus, we have an additional integration constant $f_0$.
Let us examine the implications of this new integration constant.
With $\sigma=0$, the solution is given by
\begin{equation}
\label{eq-metric-sol22}
f(r)=1+f_0,\quad
n(r)=\pm\sqrt{\frac{\xi}{\alpha}f_0
+\frac{r_{{\rm s}}}{r}-\frac{d}{r^2}}\, ,
\end{equation}
where $d=[(5\zeta+14\eta)f_0^2+2(3\zeta+8\eta)f_0]/\alpha$.
When $\xi/\alpha=1$, it looks the same as the Reissner-Nordstr\"{o}m solution
in the PG coordinates
with the identification $d=Gq^2$ where $q$ is the electric charge.
When $f_0\ne 0$, the $1/r^{2}$-term acts like that of an electric charge.
The important difference from the Reissner-Nordstr\"{o}m solution of GR is that
the coefficient of $1/r^{2}$-term can have a positive value
when $\zeta$ or $\eta<0$.
Though the quadratic curvature terms induce possibility of $d=Gq^{2}>0$,
we will read the meaning of it in the context of GR.
If we write down the Einstein equation,
$G^{\mu}_{\;\nu}=-8\pi GT^{\mu}_{\;\nu}$, in terms of the metric
\eqref{metric-sss-pg} with \eqref{f0}, it becomes
\begin{align}
(rn^{2})'=f_{0}-8\pi GT^{t}_{\; t}r^{2}= f_{0}+4\pi G f_{0}E_{r}^{2}r^{2},
\label{Eeq}
\end{align}
where the last equality holds for the radial component of electrostatic field
$E_{r}=F_{rt}$. Comparing \eqref{Eeq} with \eqref{neq}
in the limit of $\sigma=r_{{\rm s}}=\xi/\alpha -1 = 0$, we identify the
energy density as
\begin{align}
-T^{t}_{\; t}=\frac{f_{0}}{2}E_{r}^{2}
= \frac{d}{8\pi Gr^{4}}.
\end{align}
Therefore, in order to obtain the same solution $n(r)$ \eqref{eq-metric-sol22}
in GR, we need $E_{r}^{2}=q^{2}/4\pi f_{0}r^{4}$ which can be negative for
$d=Gq^{2}<0$. In the context of GR, it is forbidden since this matter
configuration violates the positive energy theorem.
In conventional electromagnetism the effect of $q^{2}<0$ may imply
sign-flipped Coulomb force, attractive between same charges.
However, in HL gravity,
the solution \eqref{eq-metric-sol22} with negative $d$ is generic.
It is obtained in the absence of matter, ${\cal L}_{{\rm M}}=0$,
but in the presence of quartic spatial derivative terms, the third term in
the right-hand side of the equation \eqref{neq}.
Since violation of the positive energy theorem and modification of such
Coulomb force are unphysical, detection of the signal of $-d/r^{2}>0$
term suggests existence of the era of HL gravity.

If $f_0$ cannot be scaled to zero in the metric of (\ref{eq-metric-sol22}),
another intriguing question is to address the effect of the constant term
$\xi f_{0}/\alpha$ which is unphysical in GR and is not related to
higher spatial derivatives.
To see this explicitly, let us examine the geodesic equation of a test body.
Let the constants of motion corresponding to the cyclic coordinates $t$
and $\phi$ be $E$ and $\ell$, respectively. Then radial geodesic equation can be written as
\begin{equation}
\dot r^2+\left(1+f_{0}\delta -\frac{r_{\rm s}}{r}+\frac{d}{r^2}\right)\left(1+\frac{\ell^2}{r^2}\right)=E^2,
\end{equation}
where the overdot denotes differentiation with respect to the proper time
of the test body.
We introduced the factor $\delta=1-\xi/\alpha$ measuring the deviation of the propagation speed from unity.
From this equation, we obtain the orbit equation for $u(\phi)\equiv1/r(\phi)$
\begin{equation}
\frac{d^{2}u}{d\phi^{2}}+\left(1+f_{0}\delta +\frac{d}{\ell^2}\right)u
  = \frac{r_{{\rm s}}}{2\ell^2}+\frac{3r_{{\rm s}}}{2}\,u^2 +2d\,u^3 .
\label{oeq}
\end{equation}
When we turn off the higher derivative terms $d=0$ and the mass
$r_{{\rm s}}=0$, the orbit equation reduces to a linear equation and
the coefficient of $u$-term decides the allowed range of the angle $\phi$.
Solution of it is $u(\phi)\propto \cos (\sqrt{1+f_{0}\delta }\,\phi)$, and
thus, non-zero $f_0$ leads to the surplus solid angle
$\Delta=2\pi(\sqrt{1+f_{0}\delta}\, -1)$ for $f_{0}\delta >0$ and
the deficit solid angle $\Delta=2\pi(1-\sqrt{1+f_{0}\delta}\, )$ for
$-1<f_{0}\delta<0$. When $f_{0}\delta=0$, the space has neither
surplus nor deficit solid angles.

In the context of GR \eqref{Eeq}, the constant metric
solution, $n=\xi f_{0}/\alpha$, from \eqref{neq} can be obtained by assuming
the energy density, $-T^{t}_{\; t}=-f_{0}\delta/8\pi Gr^{2}$.
For the deficit solid angle with $-1<f_{0}\delta<0$, the energy density is
positive and it corresponds to gravitating global
monopole~\cite{Barriola:1989hx}. For the surplus solid angle with
$f_{0}\delta>0$, the energy density is negatively distributed everywhere.
It is forbidden
in GR since it violates the positive energy theorem. However, in HL gravity,
this constant solution is a static solution attained in the absence of matter
field, ${\cal L}_{{\rm M}}=0$, and higher derivatives, $\zeta=\eta=0$.
This result is also contrasted with the case of HL gravity with the detailed
balance condition, in which a positive energy distribution of
the electrostatic field of a point charge supports the geometry involving
surplus solid angle~\cite{Kim:2009dq}.

Possible astrophysical effects of a deficit/surplus solid angle can easily be
visualized in the case
of $\theta=\pi/2$, where an observer, a light source, and the the apex of
deficit/surplus solid angle are in the same plane.
When a geometry with deficit solid angle is formed, the light from a star
behind the apex propagates straight and arrives at a static observer
who detects double images projected behind the source~\cite{Vilenkin:1981zs}.
When a geometry with surplus solid angle is formed, a static observer
tracking down the trajectory of the star experiences sudden disappearance
of its image for the period proportional to the surplus angle and
reappearance at a distant point over the apex of surplus
angle~\cite{Kim:2009dq}. The present astronomical bound of angular resolution
is about 200 micro-arcseconds ($\sim 10^{-9}$ radian)~\cite{CHA}.
It means that, in principle, it can be observed
if the graviton speed deviates away from
unity satisfies $|\delta|=|1-\xi/\alpha|>10^{-9}\pi/f_{0}$. Though $f_{0}$ is
an undetermined free parameter in the present stage, the formula suggests
$10^{-8}\sim 10^{-9}$ deviation of the graviton speed for
$f_{0}\sim {\cal O}(1)$, which looks extremely stringent for the RG flows
for the ratio $\xi/\alpha$ in the vicinity of IR fixed point.

\section{Static spherically symmetric solution
in the Poincar\'{e} coordinates}
\label{section-solution-S}

In this section, we consider static spherically symmetric vacuum solution
in non-projectable HL gravity for comparison.
This is done by taking a static spherically symmetric metric
in the Poincar\'{e} coordinates
\begin{equation}
\label{metric-sss-s}
ds^2 = -N(r)^2d{\tilde t}^2 + \frac{1}{F(r)}dr^2+r^2(d\theta^2+\sin^2\theta d\phi)^2.
\end{equation}
The metric (\ref{metric-sss-pg}) is connected to the metric
(\ref{metric-sss-s}) by a coordinate transformation
\begin{equation}
{\tilde t} = \sqrt{C}\;t + \int\frac{\sqrt{f-F}}{F}\, dr,
\end{equation}
with $r$, $\theta$, $\phi$ unchanged and
\begin{equation}
f=C\frac{F}{N^2},\qquad
n^2=(C-N^2)\frac{F}{N^2},
\end{equation}
where $C$ is a constant.
Since this transformation is not a foliation-preserving diffeomorphism,
the solutions found using the metric (\ref{metric-sss-s}) and the metric
(\ref{metric-sss-pg}) will not be equivalent in HL gravity.

The action \eqref{eq-HL-action} written in terms of $N$ and $F$ is
\begin{align}
S_{{\rm HL}} =&\ 4\pi\int dt\,dr\, \frac{r^{2}N}{\sqrt{F}}\Bigg[
  \frac{3\zeta+8\eta}{2r^{2}}{F'}^2
+ \frac{2(\zeta+4\eta)}{r^3}F'(F-1)
+ \frac{2(\zeta+2\eta)}{r^4}(F-1)^2
\nonumber\\&{}\hspace{35mm}
-\frac{2\xi}{r^{2}}(rF'+F-1)
+\sigma \Bigg],
\end{align}
and the matter-field action with spherical symmetry is
\begin{equation}
S_{{\rm M}}= \ 4\pi \int dt\,dr\, \frac{r^{2}N}{\sqrt{F}}\,
{\cal L}_{{\rm M}}(N,F,\Phi).
\end{equation}
The equations of motion obtained from the variation of $N$ and $F$ are
\begin{align}
\label{eq-sss-s-N}
& \frac{3\zeta+8\eta}{2r^2}{F'}^2
+\frac{2(\zeta+4\eta)}{r^3}F'(F-1)
+\frac{2(\zeta+2\eta)}{r^4}(F-1)^2
-\frac{2\xi}{r^2}(rF'+F-1)+\sigma
\nonumber\\ & \hspace{110mm}
=-{\cal L}_{{\rm M}}-N\frac{\partial{\cal L}_{{\rm M}}}{\partial N},
\\
\label{eq-sss-s-g}
&\left(\log\frac{N}{\sqrt{F}}\right)'
\left[ \frac{3\zeta+8\eta}{r^2}F'
+\frac{2(\zeta+4\eta)}{r^3}(F-1)-\frac{2\xi}{r} \right]
+\frac{(3\zeta+8\eta)}{r^2}\left[F''-\frac{2}{r^2}(F-1)\right]
\nonumber\\ &\hspace{110mm}
=\frac{\partial{\cal L}_{\rm M}}{\partial F}
+\frac{N}{F}\frac{\partial{\cal L}_{\rm M}}{\partial N},
\end{align}
where the prime denotes the derivative with respect to $r$.
Now we consider the vacuum solution of
(\ref{eq-sss-s-N})--(\ref{eq-sss-s-g}) for which ${\cal L}_{\rm M}=0$.
The first equation involves $F(r)$ only and
the second equation determines $N(r)$ from $F(r)$ obtained
from the first equation.
Note that both equations are independent of $\alpha$, $\beta$, and $\gamma$
since the extrinsic curvature and the Cotton tensor vanish for the
static spherically symmetric metric (\ref{metric-sss-s}).
When compared to (\ref{eq-sss-pg-N}) and (\ref{eq-sss-pg-Nr}),
a notable difference is that (\ref{eq-sss-s-N}) and (\ref{eq-sss-s-g}) are
independent of $\lambda$.

We write the metric function $F$ as
\begin{equation}
F(r)=1+r^2\left[a-p(r)\right],
\end{equation}
where
\begin{equation}
a = \frac{\xi-\tilde\xi}{4(\zeta+3\eta)}, \qquad
\tilde\xi=\sqrt{\xi^2-\frac{4}{3}\sigma(\zeta+3\eta)}\,.
\end{equation}
Then, from \eqref{eq-sss-s-N}, $p(r)$ satisfies the equation
\begin{equation}
\label{eq-p-0}
(3\zeta+8\eta)r^2{p'}^2 +4[4(\zeta+3\eta)p+\tilde\xi]rp'
+12[2(\zeta+3\eta)p^2+\tilde\xi p] = 0.
\end{equation}

When both $\zeta$ and $\eta$ vanish,
the well-known GR solution
\begin{equation}
N^2(r)=F(r)=1-\frac{\sigma}{6\xi}r^2-\frac{r_{{\rm s}}}{r}
\end{equation}
is obtained, regardless of the coefficient of scalar curvature term, $\xi$.
Here $r_{{\rm s}}$ is the same integration constant as in \eqref{nsol},
$r_{{\rm s}}=2GM$.
Note that when we have nonvanishing cosmological constant, $\sigma\ne 0$,
the asymptotic behavior is determined by the coefficient of $r^2$ term.
When higher derivative terms are introduced,
its coefficient is modified
and hence the asymptotic behavior becomes different from that of GR,
irrespective of the assignment of the detailed balance condition.

When the combination $3\zeta+8\eta$ vanishes,
the equation~(\ref{eq-p-0}) is easily integrated and the solution is given by
\begin{equation}
N^2(r)=F(r)=1+ar^2-\frac{2r^2}{\tilde\zeta}\left(
1-\sqrt{1-\frac{\tilde\zeta r_{{\rm s}}}{r^3}}\;\right),
\end{equation}
where $\tilde\zeta\equiv\zeta/\tilde\xi$.
Let us consider the case $\sigma=0$. Then, we have $a=0$ and $\tilde\xi=\xi$.
For large $r$ ($r\gg(\tilde\zeta r_{{\rm s}})^{1/3}$), we get an approximation
\begin{equation}
N^2(r)=F(r) \approx 1-\frac{r_{{\rm s}}}{r}-\frac{\tilde\zeta r_{{\rm s}}^2}{4r^4},
\end{equation}
which shows the usual behavior of the Schwarzschild black hole plus small corrections
due to the higher derivative terms.
On the other hand, at short distance, the position of event horizon is modified.
When $\tilde\zeta<0$, we have two event horizons at
\begin{equation}
r_{{\rm H}} = \frac{1}{2}\left[r_{{\rm s}}\pm\sqrt{r_{{\rm s}}^2-(-\tilde\zeta)}\right].
\end{equation}
When $\tilde\zeta>0$, we have an event horizon at
\begin{equation}
r_{{\rm H}} = \frac{1}{2}\left(r_{{\rm s}}+\sqrt{r_{{\rm s}}^2+\tilde\zeta}\right)
.
\end{equation}

For $3\zeta+8\eta\ne0$, we solve (\ref{eq-p-0}) for $p'(r)$ and obtain
\begin{equation}
\label{eq-p-1}
\tilde p'=-\frac{6}{(1+8b)r} \left(1+4b\tilde p-\sqrt{1-\tilde p-2b\tilde
p^2}\ \right),
\end{equation}
where $\tilde p=\tilde\zeta p$ and $b=1+3\eta/\zeta$.
When $b=0$, this equation is easily integrated to give
\begin{equation}
(\sqrt{1-\tilde p}-1)e^{(\sqrt{1-\tilde p}-1)} = -\frac{\tilde\zeta r_{{\rm s}}}{2r^3}.
\end{equation}
For $b\ne0$, we get
\begin{equation}
\frac{\tilde p^2\left[1+2b\left(2+\tilde p+2\sqrt{1-\tilde p-2b\tilde p^2}
\right)\right]
\left(\frac{1+4b\tilde p}{\sqrt{-2b}}+2\sqrt{1-\tilde p-2b\tilde p^2}\right)^{\frac{1}{\sqrt{-2b}}}}
{2-\tilde p+2\sqrt{1-\tilde p-2b\tilde p^2}}=\frac{c}{r^6},
\end{equation}
where $c$ is a constant.
To get $\tilde p(r)$ explicitly, we need to invert these equations.
Assuming $\tilde p$ is small, we can solve the equation perturbatively and obtain
\begin{equation}
\tilde p \approx -\frac{\tilde\zeta r_{{\rm s}}}{r^3} \left(1-\frac{\tilde\zeta r_{{\rm s}}}{4r^3}\right)^{-1}.
\end{equation}
Then up to this order
\begin{eqnarray}
F(r) &=& 1+ar^2-\frac{r_{{\rm s}}}{r}-\frac{\zeta r_{{\rm s}}^2}{4\tilde\xi r^4}, \\
N^2(r) &=& F(r) + \frac{3a\zeta(3\zeta+8\eta)r_{{\rm s}}^2}{4\tilde\xi^2r^4}.
\end{eqnarray}
In general, the higher derivative terms give rise to the subleading corrections of order
$\zeta r_{{\rm s}}/\xi r^3$. Thus, Eddington-Robertson parameters are same
as those of GR, and it is hard to observe the macroscopic effect of them.
However, higher spatial derivative terms modify the causal structure at short distance.

In the previous and present sections, we obtain static vacuum solution of a
projectable and a non-projectable version of HL gravity without the detailed
balance condition at the IR regime, which are differentiated by two
inequivalent metrics. For vanishing cosmological constant case, static
spherically symmetric solution for the metric without nontrivial lapse function
in the PG coordinates supports geometry of a surplus or deficit solid angle,
however the other solution in the Poincar\'{e} coordinates does not.
Instead of such a long range effect, it involves a short distance correction
like the change of event horizons.

\section{Conclusion}
\label{section-conclusion}

We obtained static spherically symmetric vacuum solutions
of the recently proposed HL gravity,
imposing the detailed balance condition only in the UV limit.
Since HL gravity is intrinsically non-relativistic,
the choice of a preferred frame is unavoidable.
The choice of coordinate system is related to the choice of the frame.
For static spherically symmetric vacuum solutions,
we tried two coordinate systems, the PG coordinates and
the Poincar\'{e} coordinates.
They are connected by the time-space mixing coordinate transformation
which is not foliation-preserving.
Thus, the solutions in two coordinate systems are distinct and
physically inequivalent.
The distinguishing feature of metrics in two coordinate systems is that
the one in PG coordinates satisfies the projectability condition and
the other does not.

Even in the absence of higher derivative terms,
the solution without matter distribution in the PG coordinates has an additional
integration constant which leads to the geometry of a surplus or deficit
solid angle. This long range effect can be constrained by smallness
of the astronomical bound of angular resolution, and so do some parameters
of HL gravity.
In both solutions, the effect of higher spatial derivative terms is that
they change the causal structures at short distances.
But the differences from GR are not observationally significant
in the macroscopic world for the reasonable choice of parameters.
This is as expected because the higher spatial derivative terms are
in general suppressed by the Planck scale or, if any, the new quantum
gravity scale which may be smaller than the Planck scale but still out
of our experimental reach.

The solution in PG coordinate system may be stated as
the static spherically symmetric vacuum solution of projectable HL gravity.
For this solution we found that
a new integration constant must be introduced in addition to mass,
when $\xi/\alpha$ differs from unity or we have non-vanishing higher
spatial derivative terms.
This constant implies the existence of surplus or deficit angle for the former
and the electric charge for the latter.
The current astronomical bound on surplus/deficit angle can impose stringent
constraint on the deviation of $\xi/\alpha$ from unity.
One striking feature of the solution is that the electric charge can
effectively be imaginary depending on the signature of the couplings of higher
spatial derivative terms.

\section*{Acknowledgments}

We would like to thank Taekyung Kim for valuable discussions.
This work was supported by the Science Research Center Program of the
Korean Science and Engineering Foundation (KOSEF)
through the Center for Quantum SpaceTime(CQUeST) of Sogang University
with grant number R11-2005-021,
and by the Korea Science and Engineering Foundation (KOSEF) grant
funded by the Korea government (MEST) (No.~2009-0075127) (HBK).
This work was also supported by Astrophysical Research Center for the
Structure and Evolution of the Cosmos (ARCSEC)) and
by the Korea Research Foundation Grant funded by the Korean Government
(KRF-2008-313-C00170) (YK).


\begin{thebibliography}{0}

\bibitem{Horava:2009uw}
  P.~Horava,
  ``Quantum Gravity at a Lifshitz Point,''
  Phys.\ Rev.\  D {\bf 79}, 084008 (2009)
  [arXiv:0901.3775 [hep-th]].

\bibitem{Horava:2010zj}
  P.~Horava and C.~M.~Melby-Thompson,
  ``General Covariance in Quantum Gravity at a Lifshitz Point,''
  arXiv:1007.2410 [hep-th].


\bibitem{Visser:2009fg}
  M.~Visser,
  ``Lorentz symmetry breaking as a quantum field theory regulator,''
  Phys.\ Rev.\  D {\bf 80}, 025011 (2009)
  [arXiv:0902.0590 [hep-th]];
%
  H.~Nikolic,
  ``Horava-Lifshitz gravity, absolute time, and objective particles in curved
  space,''
  Mod.\ Phys.\ Lett.\  A {\bf 25}, 1595 (2010)
  [arXiv:0904.3412 [hep-th]];
%
  G.~E.~Volovik,
  ``z=3 Lifshitz-Horava model and Fermi-point scenario of emergent gravity,''
  JETP Lett.\  {\bf 89}, 525 (2009)
  [arXiv:0904.4113 [gr-qc]];
%
  B.~Chen and Q.~G.~Huang,
  ``Field Theory at a Lifshitz Point,''
  Phys.\ Lett.\  B {\bf 683}, 108 (2010)
  [arXiv:0904.4565 [hep-th]];
%
  R.~G.~Cai, B.~Hu and H.~B.~Zhang,
  ``Dynamical Scalar Degree of Freedom in Horava-Lifshitz Gravity,''
  Phys.\ Rev.\  D {\bf 80}, 041501 (2009)
  [arXiv:0905.0255 [hep-th]];
%
  D.~Orlando and S.~Reffert,
  ``On the Renormalizability of Horava-Lifshitz-type Gravities,''
  Class.\ Quant.\ Grav.\  {\bf 26}, 155021 (2009)
  [arXiv:0905.0301 [hep-th]];
%
  C.~Gao,
  ``Modified gravity in Arnowitt-Deser-Misner formalism,''
  Phys.\ Lett.\  B {\bf 684}, 85 (2010)
  [arXiv:0905.0310 [astro-ph.CO]];
%
  T.~Nishioka,
  ``Horava-Lifshitz Holography,''
  Class.\ Quant.\ Grav.\  {\bf 26}, 242001 (2009)
  [arXiv:0905.0473 [hep-th]];
%
  J.~Chen and Y.~Wang,
  ``Timelike Geodesic Motion in Horava-Lifshitz Spacetime,''
  Int.\ J.\ Mod.\ Phys.\  A {\bf 25}, 1439 (2010)
  [arXiv:0905.2786 [gr-qc]];
%
  G.~Calcagni,
  ``Detailed balance in Horava-Lifshitz gravity,''
  Phys.\ Rev.\  D {\bf 81}, 044006 (2010)
  [arXiv:0905.3740 [hep-th]];
%
  Y.~S.~Myung,
  ``Propagations of massive graviton in the deformed Ho\v{r}ava-Lifshitz gravity,''
  Phys.\ Rev.\  D {\bf 81}, 064006 (2010)
  [arXiv:0906.0848 [hep-th]];
%
  C.~Germani, A.~Kehagias and K.~Sfetsos,
  ``Relativistic Quantum Gravity at a Lifshitz Point,''
  JHEP {\bf 0909}, 060 (2009)
  [arXiv:0906.1201 [hep-th]];
%
  F.~W.~Shu and Y.~S.~Wu,
  ``Stochastic Quantization of the Ho\v{r}ava Gravity,''
  arXiv:0906.1645 [hep-th];
%
  R.~Iengo, J.~G.~Russo and M.~Serone,
  ``Renormalization group in Lifshitz-type theories,''
  JHEP {\bf 0911}, 020 (2009)
  [arXiv:0906.3477 [hep-th]];
%
  D.~Blas, O.~Pujolas and S.~Sibiryakov,
  ``Consistent Extension Of Horava Gravity,''
  Phys.\ Rev.\ Lett.\  {\bf 104}, 181302 (2010)
  [arXiv:0909.3525 [hep-th]];
%
  J.~Bellorin and A.~Restuccia,
  ``On the consistency of the Horava Theory,''
  arXiv:1004.0055 [hep-th];
%
  J.~M.~Pons and P.~Talavera,
  ``Remarks on the consistency of minimal deviations from General Relativity,''
  Phys.\ Rev.\  D {\bf 82}, 044011 (2010)
  [arXiv:1003.3811 [gr-qc]].

\bibitem{Charmousis:2009tc}
  C.~Charmousis, G.~Niz, A.~Padilla and P.~M.~Saffin,
  ``Strong coupling in Horava gravity,''
  JHEP {\bf 0908}, 070 (2009)
  [arXiv:0905.2579 [hep-th]];
%
  D.~Blas, O.~Pujolas and S.~Sibiryakov,
  ``On the Extra Mode and Inconsistency of Horava Gravity,''
  JHEP {\bf 0910}, 029 (2009)
  [arXiv:0906.3046 [hep-th]].

\bibitem{Sotiriou:2009bx}
  T.~P.~Sotiriou, M.~Visser and S.~Weinfurtner,
  ``Quantum gravity without Lorentz invariance,''
  JHEP {\bf 0910}, 033 (2009)
  [arXiv:0905.2798 [hep-th]].

\bibitem{Kim:2009zn}
  Y.~W.~Kim, H.~W.~Lee and Y.~S.~Myung,
  ``Nonpropagation of scalar in the deformed Ho\v{r}ava-Lifshitz gravity,''
  Phys.\ Lett.\  B {\bf 682}, 246 (2009)
  [arXiv:0905.3423 [hep-th]].

\bibitem{Li:2009bg}
  M.~Li and Y.~Pang,
  ``A Trouble with Ho\v{r}ava-Lifshitz Gravity,''
  JHEP {\bf 0908}, 015 (2009)
  [arXiv:0905.2751 [hep-th]];
%
  M.~Henneaux, A.~Kleinschmidt and G.~L.~Gomez,
  ``A dynamical inconsistency of Horava gravity,''
  Phys.\ Rev.\  D {\bf 81}, 064002 (2010)
  [arXiv:0912.0399 [hep-th]].


\bibitem{Lu:2009em}
  H.~Lu, J.~Mei and C.~N.~Pope,
  ``Solutions to Horava Gravity,''
  Phys.\ Rev.\ Lett.\  {\bf 103}, 091301 (2009)
  [arXiv:0904.1595 [hep-th]].

\bibitem{Nastase:2009nk}
  H.~Nastase,
  ``On IR solutions in Horava gravity theories,''
  arXiv:0904.3604 [hep-th];
%
  R.~G.~Cai, L.~M.~Cao and N.~Ohta,
  ``Topological Black Holes in Horava-Lifshitz Gravity,''
  Phys.\ Rev.\  D {\bf 80}, 024003 (2009)
  [arXiv:0904.3670 [hep-th]];
%
  E.~O.~Colgain and H.~Yavartanoo,
  ``Dyonic solution of Horava-Lifshitz Gravity,''
  JHEP {\bf 0908}, 021 (2009)
  [arXiv:0904.4357 [hep-th]];
%
  Y.~S.~Myung and Y.~W.~Kim,
  ``Thermodynamics of Ho\v{r}ava-Lifshitz black holes,''
  Eur.\ Phys.\ J.\  C {\bf 68}, 265 (2010)
  [arXiv:0905.0179 [hep-th]];
%
  A.~Kehagias and K.~Sfetsos,
  ``The black hole and FRW geometries of non-relativistic gravity,''
  Phys.\ Lett.\  B {\bf 678}, 123 (2009)
  [arXiv:0905.0477 [hep-th]];
%
  R.~G.~Cai, L.~M.~Cao and N.~Ohta,
  ``Thermodynamics of Black Holes in Horava-Lifshitz Gravity,''
  Phys.\ Lett.\  B {\bf 679}, 504 (2009)
  [arXiv:0905.0751 [hep-th]];
%
  A.~Ghodsi,
  ``Toroidal solutions in Horava Gravity,''
  arXiv:0905.0836 [hep-th];
%
  Y.~S.~Myung,
  ``Thermodynamics of black holes in the deformed Ho\v{r}ava-Lifshitz gravity,''
  Phys.\ Lett.\  B {\bf 678}, 127 (2009)
  [arXiv:0905.0957 [hep-th]];
%
  S.~Chen and J.~Jing,
  ``Quasinormal modes of a black hole in the deformed Ho\v{r}ava-Lifshitz gravity,''
  Phys.\ Lett.\  B {\bf 687}, 124 (2010)
  [arXiv:0905.1409 [gr-qc]];
%
  D.~W.~Pang,
  ``A Note on Black Holes in Asymptotically Lifshitz Spacetime,''
  arXiv:0905.2678 [hep-th];
%
  G.~Bertoldi, B.~A.~Burrington and A.~Peet,
  ``Black Holes in asymptotically Lifshitz spacetimes with arbitrary critical exponent,''
  Phys.\ Rev.\  D {\bf 80}, 126003 (2009)
  [arXiv:0905.3183 [hep-th]];
%
  M.~i.~Park,
  ``The Black Hole and Cosmological Solutions in IR modified Horava Gravity,''
  JHEP {\bf 0909}, 123 (2009)
  [arXiv:0905.4480 [hep-th]];
%
  M.~Botta-Cantcheff, N.~Grandi and M.~Sturla,
  ``Wormhole solutions to Horava gravity,''
  arXiv:0906.0582 [hep-th];
%
  A.~Ghodsi and E.~Hatefi,
  ``Extremal rotating solutions in Horava Gravity,''
  Phys.\ Rev.\  D {\bf 81}, 044016 (2010)
  [arXiv:0906.1237 [hep-th]];
%
  A.~Castillo and A.~Larranaga,
  ``Entropy for Black Holes in the Deformed Horava-Lifshitz Gravity,''
  arXiv:0906.4380 [gr-qc];
%
  H.~W.~Lee, Y.~W.~Kim and Y.~S.~Myung,
  ``Extremal black holes in the Ho\v{r}ava-Lifshitz gravity,''
  Eur.\ Phys.\ J.\  C {\bf 68}, 255 (2010)
  [arXiv:0907.3568 [hep-th]];
%
  J.~Z.~Tang and B.~Chen,
  ``Static Spherically Symmetric Solutions to modified Horava-Lifshitz Gravity with Projectability Condition,''
  Phys.\ Rev.\  D {\bf 81}, 043515 (2010)
  [arXiv:0909.4127 [hep-th]];
%
  E.~Kiritsis and G.~Kofinas,
  ``On Horava-Lifshitz 'Black Holes',''
  JHEP {\bf 1001}, 122 (2010)
  [arXiv:0910.5487 [hep-th]];
%
  D.~Capasso and A.~P.~Polychronakos,
  ``General static spherically symmetric solutions in Horava gravity,''
  Phys.\ Rev.\  D {\bf 81}, 084009 (2010)
  [arXiv:0911.1535 [hep-th]];
%
  B.~R.~Majhi,
  ``Hawking radiation and black hole spectroscopy in Horava-Lifshitz gravity,''
  Phys.\ Lett.\  B {\bf 686}, 49 (2010)
  [arXiv:0911.3239 [hep-th]];
%
  J.~Z.~Tang,
  ``Static Charged Black Hole Solutions in Horava-Lifshitz Gravity,''
  arXiv:0911.3849 [hep-th].
%
  E.~Gruss,
  ``Black Holes in Ho\v{r}ava Gravity with Higher Derivative Magnetic Terms,''
  arXiv:1005.1353 [hep-th].


\bibitem{Calcagni:2009ar}
  G.~Calcagni,
  ``Cosmology of the Lifshitz universe,''
  JHEP {\bf 0909}, 112 (2009)
  [arXiv:0904.0829 [hep-th]];
%
  R.~Brandenberger,
  ``Matter Bounce in Horava-Lifshitz Cosmology,''
  Phys.\ Rev.\  D {\bf 80}, 043516 (2009)
  [arXiv:0904.2835 [hep-th]];
%
  S.~Kalyana Rama,
  ``Anisotropic Cosmology and (Super)Stiff Matter in Ho\v{r}ava's Gravity Theory,''
  Phys.\ Rev.\  D {\bf 79}, 124031 (2009)
  [arXiv:0905.0700 [hep-th]];
%
  E.~N.~Saridakis,
  ``Horava-Lifshitz Dark Energy,''
  Eur.\ Phys.\ J.\  C {\bf 67}, 229 (2010)
  [arXiv:0905.3532 [hep-th]];
%
  S.~Mukohyama,
  ``Dark matter as integration constant in Horava-Lifshitz gravity,''
  Phys.\ Rev.\  D {\bf 80}, 064005 (2009)
  [arXiv:0905.3563 [hep-th]];
%
  M.~Minamitsuji,
  ``Classification of cosmology with arbitrary matter in the Ho\v{r}ava-Lifshitz theory,''
  Phys.\ Lett.\  B {\bf 684}, 194 (2010)
  [arXiv:0905.3892 [astro-ph.CO]];
%
  A.~Wang and Y.~Wu,
  ``Thermodynamics and classification of cosmological models in the Horava-Lifshitz theory of gravity,''
  JCAP {\bf 0907}, 012 (2009)
  [arXiv:0905.4117 [hep-th]];
%
  S.~Nojiri and S.~D.~Odintsov,
  ``Covariant Horava-like renormalizable gravity and its FRW cosmology,''
  Phys.\ Rev.\  D {\bf 81}, 043001 (2010)
  [arXiv:0905.4213 [hep-th]].

\bibitem{Kiritsis:2009sh}
  E.~Kiritsis and G.~Kofinas,
  ``Horava-Lifshitz Cosmology,''
  Nucl.\ Phys.\  B {\bf 821}, 467 (2009)
  [arXiv:0904.1334 [hep-th]].


\bibitem{Takahashi:2009wc}
  T.~Takahashi and J.~Soda,
  ``Chiral Primordial Gravitational Waves from a Lifshitz Point,''
  Phys.\ Rev.\ Lett.\  {\bf 102}, 231301 (2009)
  [arXiv:0904.0554 [hep-th]];
%
  S.~Mukohyama,
  ``Scale-invariant cosmological perturbations from Horava-Lifshitz gravity without inflation,''
  JCAP {\bf 0906}, 001 (2009)
  [arXiv:0904.2190 [hep-th]];
%
  Y.~S.~Piao,
  ``Primordial Perturbation in Horava-Lifshitz Cosmology,''
  Phys.\ Lett.\  B {\bf 681}, 1 (2009)
  [arXiv:0904.4117 [hep-th]];
%
  X.~Gao,
  ``Cosmological Perturbations and Non-Gaussianities in Ho\v{r}ava-Lifshitz Gravity,''
  arXiv:0904.4187 [hep-th];
%
  B.~Chen, S.~Pi and J.~Z.~Tang,
  ``Scale Invariant Power Spectrum in Ho\v{r}ava-Lifshitz Cosmology without Matter,''
  JCAP {\bf 0908}, 007 (2009)
  [arXiv:0905.2300 [hep-th]];
%
  X.~Gao, Y.~Wang, R.~Brandenberger and A.~Riotto,
  ``Cosmological Perturbations in Ho\v{r}ava-Lifshitz Gravity,''
  Phys.\ Rev.\  D {\bf 81}, 083508 (2010)
  [arXiv:0905.3821 [hep-th]].



\bibitem{Sotiriou:2009gy}
  T.~P.~Sotiriou, M.~Visser and S.~Weinfurtner,
  ``Phenomenologically viable Lorentz-violating quantum gravity,''
  Phys.\ Rev.\ Lett.\  {\bf 102}, 251601 (2009)
  [arXiv:0904.4464 [hep-th]];
%
  S.~Mukohyama, K.~Nakayama, F.~Takahashi and S.~Yokoyama,
  ``Phenomenological Aspects of Horava-Lifshitz Cosmology,''
  Phys.\ Lett.\  B {\bf 679}, 6 (2009)
  [arXiv:0905.0055 [hep-th]];
%
  R.~A.~Konoplya,
  ``Towards constraining of the Horava-Lifshitz gravities,''
  Phys.\ Lett.\  B {\bf 679}, 499 (2009)
  [arXiv:0905.1523 [hep-th]];
%
  S.~b.~Chen and J.~l.~Jing,
  ``Strong field gravitational lensing in the deformed H\v{o}rava-Lifshitz black hole,''
  Phys.\ Rev.\  D {\bf 80}, 024036 (2009)
  [arXiv:0905.2055 [gr-qc]];
%
  N.~Afshordi,
  ``Cuscuton and low energy limit of Horava-Lifshitz gravity,''
  Phys.\ Rev.\  D {\bf 80}, 081502 (2009)
  [arXiv:0907.5201 [hep-th]];
%
  S.~Dutta and E.~N.~Saridakis,
  ``Overall observational constraints on the running parameter $\lambda$ of Horava-Lifshitz gravity,''
  JCAP {\bf 1005}, 013 (2010)
  [arXiv:1002.3373 [hep-th]].


\bibitem{Barriola:1989hx}
  M.~Barriola and A.~Vilenkin,
  ``Gravitational Field of a Global Monopole,''
  Phys.\ Rev.\ Lett.\  {\bf 63}, 341 (1989).

\bibitem{Kim:2009dq}
  S.~S.~Kim, T.~Kim and Y.~Kim,
  ``Surplus Solid Angle: Toward Astrophysical Test of Horava-Lifshitz Gravity,''
  Phys.\ Rev.\  D {\bf 80}, 124002 (2009)
  [arXiv:0907.3093 [hep-th]];
  ``Surplus Solid Angle in Horava-Lifshitz
  Gravity,''
  J. Korean Phys. Soc. {\bf 57}, 634 (2010).

\bibitem{Vilenkin:1981zs}
  A.~Vilenkin,
  ``Gravitational field of vacuum domain walls and strings,''
  Phys.\ Rev.\ D {\bf 23}, 852 (1981).
For a review, see A. Vilenkin and E.~P.~S.~Shellard,
{\it Cosmic strings and other topological defects}, (Cambridge
University Press, 1984).

\bibitem{CHA}
Refer to the web site of High Angular Resolution Astronomy (CHARA)
designed for exceptionally high angular resolution:
http://www.chara.gsu.edu/CHARA/
It is optical/interferrometric array of six telescopes and its individual
telescope operates at visible and near IR wavelengths.

\end{thebibliography}
\end{document}